\documentclass[12pt]{article}
\usepackage{bm}        

\usepackage{authblk}
\usepackage{float}
\usepackage{latexsym,amsmath,amscd,amssymb,graphics}
\usepackage{enumerate}
\usepackage{wrapfig}
\usepackage{graphicx}
\usepackage{framed}
\usepackage{url}
\usepackage{color}
\usepackage{amsthm}
\usepackage{cancel}
\usepackage[utf8]{inputenc} 
\usepackage{enumitem} 
\usepackage{cite}

\newcommand{\norm}[1]{\left\lVert#1\right\rVert}

\newcommand{\aaa}[1]{\left \langle#1\right\rangle}

\newcommand{\pp}[2]{\frac{\partial #1}{\partial #2}}

\newcommand{\bR}{\boldsymbol{R}}
\newcommand{\bO}{\boldsymbol{\Omega}}
\newcommand{\bG}{\boldsymbol{\Gamma}}
\newcommand{\bA}{\boldsymbol{A}}
\newcommand{\br}{\boldsymbol{r}}

\newcommand{\balpha}{\boldsymbol{\alpha}}
\newcommand{\bS}{\boldsymbol{\Sigma}}

\newcommand{\bT}{\boldsymbol{T}}
\newcommand{\bN}{\boldsymbol{N}}

\newcommand{\bPi}{\boldsymbol{\Pi}}

\newcommand{\bP}{\boldsymbol{P}}
\newcommand{\mbP}{\mathbf{P}}
\newcommand{\bp}{\boldsymbol{\psi}}
\newcommand{\bE}{\boldsymbol{E}}
\newcommand{\be}{\boldsymbol{e}}
\newcommand{\bY}{\boldsymbol{Y}}
\newcommand{\bX}{\boldsymbol{X}}

\newcommand{\bI}{\mathbb{I}}

\newcommand{\Om}{\Omega}
\newcommand{\al}{\boldsymbol{\alpha}}

\newtheorem{theorem}{Theorem}

\newtheorem{remark}[theorem]{Remark}

\newcommand{\todo}[1]{\vspace{5 mm}\par \noindent
\framebox{\begin{minipage}[c]{0.95 \textwidth}
\tt #1 \end{minipage}}\vspace{5 mm}\par}
\newcommand{\rem}[1]{} 
\DeclareMathOperator{\arctanh}{arctanh}

\newcommand{\mse}{\mathfrak{se}}

\begin{document}

\title{Integrability and Chaos in Figure Skating}
\author{Vaughn Gzenda$^1$ and Vakhtang Putkaradze$^2$ \\  
Email: $^1$gzenda@ualberta.ca, $^2$putkarad@ualberta.ca}

\affil{Department of Mathematical and Statistical Sciences, University of Alberta, \\
Edmonton, AB T6G 1S5 Canada 
}
\date{\today}
\maketitle
\begin{abstract}

We derive and analyze a three dimensional model of a figure skater. We model the skater as a three-dimensional body moving in space subject to a non-holonomic constraint enforcing movement along the skate's direction and holonomic constraints of continuous contact with ice and pitch constancy of the skate. For a static (non-articulated) skater, we show that the system is integrable  if and only if the projection of the center of mass on skate's direction coincides with the contact point with ice and some mild (and realistic) assumptions on the directions of inertia's axes. The integrability is proved by showing the existence of two new constants of motion linear in momenta, providing a new and highly nontrivial example of an integrable non-holonomic mechanical system. We also consider the case when the projection of the center of mass on skate's direction does not coincide with the contact point and show that  this  non-integrable case  exhibits  apparent chaotic behavior, by studying the divergence of nearby trajectories.  We also demonstrate the intricate behavior during the transition from the integrable to chaotic case. Our model shows many features of real-life skating, especially figure skating, and we conjecture that real-life skaters may intuitively use the discovered mechanical properties of the system for the control of the performance on ice. 
\end{abstract}

\maketitle

\section{Introduction.} 
Figure skating is a beautiful and  popular sport combining elegance, athleticism and precision, in a seemingly effortless and artful performance. The physics of skating itself, \emph{i.e.}, the description of a blade sliding on ice, has attracted considerable attention, with research focusing on the physics of ice melting under the blade and resulting friction forces \cite{Ro2005,LoSzMa2013,PoLeBe2015}. The two-dimensional model of a skater has been a highly popular topic in the literature on mechanics. This system,  the so-called Chaplygin's sleigh,  represents a flat object which can move without friction on ice along the direction of a blade chosen by the orientation of the body \cite{Bloch2003,NeFu1972}. It is a model for a sled, or alternatively, a skater that is additionally supported by frictionless legs preventing any tilting.  Chaplygin's sleigh was shown to provide rich structure of regular and chaotic behavior \cite{BlMaZe2009,BiBoMa2017,BiBoKoMa2018} and  forms one of the favourite cases of study of a system with non-holonomic  constraints  \cite{NeFu1972,Bloch2003,Ho2011_pII,deLe2012,BoMaBi2016}. However, as far as we are aware, there have been no studies of the three-dimensional motion of a skater using the modern tools of non-holonomic mechanics. 

We treat the skate as an object which slides without friction along the blade's direction, but cannot move normal to the blade or detach from ice.   This model of the skate's motion, incorporated into the fully three-dimensional motion of a skater, leads to an example of a non-holonomic system.  We show the surprising result that in spite of the apparent high complexity of the system describing the three-dimensional motion of the skater and  trajectories on ice, the system is integrable  when the center of mass is exactly balanced in the direction along to the skate, while being arbitrarily unbalanced in the sideways direction. If the center of mass is moved either forward or backward with respect to the skate, the motion becomes chaotic. Integrable non-holonomic systems are exceedingly rare, with only a handful of examples available \cite{BaCu1999,Bloch2003,Ko2002} and thus the integrability in this highly complex system is intriguing and non-intuitive.

The paper is structured as follows. In Section~\ref{sec:setup}, we present the setup-up of the system, introduce the coordinate frames and constraints, both holonomic and non-holonomic, and derive the equations of motion. In 
Section~\ref{sec:integrable}, we derive the necessary and sufficient condition of integrability due to the presence of integrals of motion linear in the momenta. In Section~\ref{sec:chaotic}, we numerically investigate the integrable and non-integrable cases, and show that the numerical evidence points to the chaotic behavior of the non-integrable case due to the exponential divergence of the nearby trajectories. Finally, Section~\ref{sec:conclusions} draws conclusions from the paper and poses some interesting directions for further studies. 

\section{Set-up of coordinates, variables and equations of motion} 
\label{sec:setup} 
\subsection{Variables and constraints} 
\begin{figure}[H]
\centering 
\includegraphics[width=0.8\textwidth]{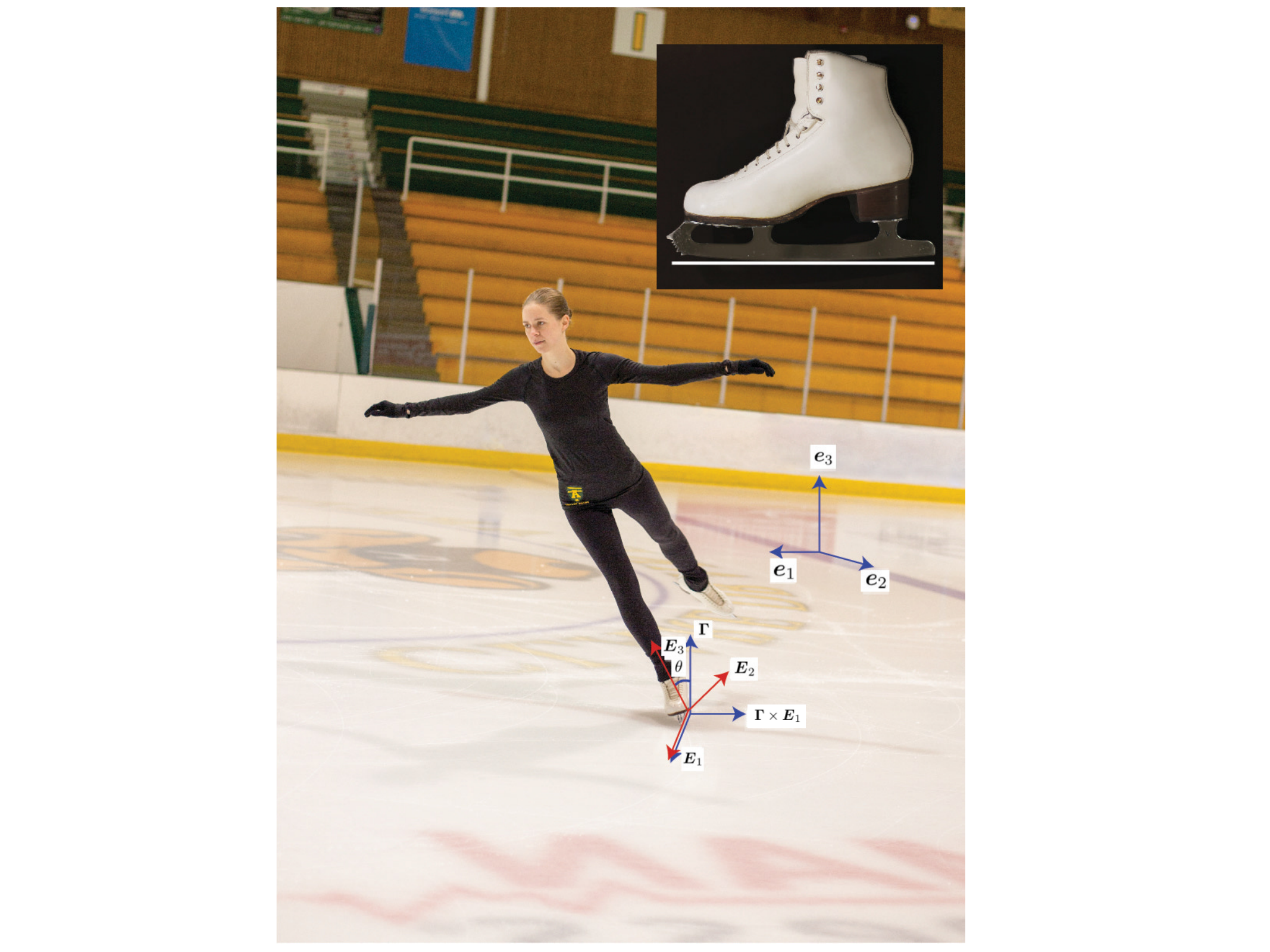} 
\caption{An Illustration of a skater with coordinate axes, as described in  Section~\ref{sec:setup}. The spatial (laboratory) frame is $\{ \mathbf{e}_1,\mathbf{e}_2,\mathbf{e}_3 \}$ and the frame attached to the boot is $\{ \mathbf{E}_1,\mathbf{E}_2,\mathbf{E}_3 \}$. The skate's blade is aligned with vector $\mathbf{E}_1$, the vector $\bG$, normal to $\mathbf{E}_1$ indicates the vertical direction as seen from the skate's frame. The vector $\mathbf{E}_1 \times \bG$ determines the direction normal to the skate and parallel to the ice. Insert: a picture of a figure skating boot and blade, illustrating a slight curvature of the blade enforcing a point contact with ice and free gliding and rotation, at the same time preventing the change of the pitch in the forward or backward direction. 
 \label{fig:skater_with_axis}
}  
\end{figure}

A skater's motion is due to the effects of inertia, gravity,  limb motion, reaction from ice on the blade and friction. In order to describe the motion of a skater on ice, as shown on Figure~\ref{fig:skater_with_axis}, we model a skater to be a possibly flexible body moving on an idealized skate, where a large force of friction prevents the skate from moving normal to its direction and there is no friction associated with the motion along the blade.  

The notations for description of a skater are illustrated on Figure~\ref{fig:skater_with_axis}. We describe the skater by the position of skate on ice $\br = (r_1,r_2,0)$ with coordinates being functions of time. We consider the skate as a rigid body  with the axes of the skate's frame given by $\{ \mathbf{E}_1, \mathbf{E}_2, \mathbf{E}_3\}$, the axes of the spatial frame are $\{ \mathbf{e}_1, \mathbf{e}_2, \mathbf{e}_3\}$ and $\Lambda$ being the transformation matrix between the spatial (laboratory) and skate (body) frames, Thus, the configuration manifold of the system is $SE(3)=SO(3) \ltimes \mathbb{R}^3$, the group of rotations and translations, with the additional constraint stating that the skate does not leave the ice, \emph{i.e.}, $r_3=\aaa{\mathbf{r},\mathbf{e}_3}=0$. We could also have chosen the configuration manifold to be $SO(3) \times \mathbb{R}^2$, treating $\mathbf{r}$ as a two-dimensional vector. While these descriptions are equivalent, we prefer the $SE(3)$ description with the additional constraint as it allows better utilization of the group structure and is more convenient for the hybrid frame description outlined in Section~\ref{sec:hybrid_frame}. 

It is important to empasize that it is only the motion of the skate that is described by the rotations and translations, while the skater may  articulate  parts of the body with respect to the skate. We shall only treat the dynamics when the skater's position is static, \emph{i.e.}, the body parts are not moving with respect to the body itself. The equations of motion for a body with moveable limbs can be derived using the theory of pseudo-rigid bodies \cite{HoSchSt2009}, Ch.10 and the Lagrange-d'Alembert's principle for non-holonomic constraints  which  will be treated in future work. In this paper, we set all the parameters of the skater in body variables to be independent of time.  

  We will present our work exclusively in the skate's frame $\{ \mathbf{E}_1, \mathbf{E}_2, \mathbf{E}_3\}$ as it is the most natural setting for describing this physical system: for the case of a non-articulated skater, the kinetic energy is left-invariant with respect to rotations and translations.   We define the following variables: 
\begin{itemize}
\setlength{\itemsep}{0pt} 
\item $\bY = \Lambda^T\dot{\br}$, the linear velocity of the skate.
\item$\bR = \Lambda^T \br$, the position of the point of contact of the skate with the ice.
\item $\bA$, the position of the centre of mass in skate's frame. 
\item $\bG=\Lambda^T \mathbf{e}_3$, the vertical axis as seen from the skate's frame. 
\end{itemize}
As can be seen by taking a time derivative, $\bG=\Lambda^T \mathbf{e}_3$  satisfies $\dot \bG = - \bO \times \bG$. The position of the center of mass in the spatial frame is then given by $\br + \Lambda \bA$, and the velocity in the spatial frame is $\mathbf{v}_{CM}=\dot \br + \dot \Lambda \bA + \Lambda \dot \bA$. The velocity of the center of mass in the skate's frame is then $\mathbf{V}_{CM}=\Lambda^T \mathbf{v}_{CM}=\bY + \bO \times \bA + \dot \bA$.  In what follows, we assume that the skater is static (non-actuated), and put $\dot \bA=\mathbf{0}$.   If the mass of the skater is $m$, moment of inertia about the center of mass $\mathbb{I} $, the Lagrangian of the system, defined as the difference between the kinetic energy and potential energy, is computed as: 
\begin{equation}
\label{Lagr_body} 
L = \frac{1}{2}\aaa{\mathbb{I}\bO , \bO }+  \frac{1}{2}m \norm{\bO \times \bA  + \bY}^2  - mg \aaa{ \bA ,\bG}.
\end{equation}
The first term is the kinetic energy of rotation about the center of mass, which we have expressed in terms of the angular velocity $\bO$ measured in the body frame,  $\aaa{\mathbf{a}, \mathbf{b}}$ is the  Euclidean  scalar product of two vectors. The second term is the kinetic energy of the linear motion of the  center of mass, and the last term is minus the potential energy of the center of mass due to gravity, with all variables measured in the skate frame $\{\mathbf{E}_1,\mathbf{E}_2,\mathbf{E}_3\}$.

Next, the set of constraints satisfied by the system are: 
\begin{enumerate} 
\setlength{\itemsep}{0pt} 
\item 
\emph{Pitch constancy}. The blade of the skate cannot tilt forward or backward with respect to the ice. In other words, the blade cannot 'dig into ice' with either the toe or the heel.  This is known as pitch constancy.  An insert in Figure~\ref{fig:skater_with_axis} illustrates a slight curvature of the figure skating blade, making this assumption correct to within a few degrees. This constraint is formulated in the skate's variables as 
\begin{equation}
\label{no_tilt} 
\aaa{\bE_1,\bG} = 0.
\end{equation}
Thus, the skate always remains normal to the vertical, whether seen in the skate's or spatial frame. We can parameterize $\bG$ by a single inclination angle $\theta$ as on Figure~\ref{fig:skater_with_axis} as $\bG=(0,\sin\theta,\cos\theta)$. We shall remark that $\theta=\pm \pi/2$ corresponds to the skate being parallel to the ice, which is technically a singularity that we do not consider in our model. 
\item 
\emph{Continuous contact}. The skate must always stay in the plane of the ice, which we can write in the spatial frame as
$\aaa{\br, \be_3} = 0$.  By multiplication on the left by $\Lambda^T$ we can rewrite this (holonomic) constraint in the skate's frame variables
\begin{equation}
\label{no_lift} 
\aaa{\bR , \bG} = 0.
\end{equation} 
\item 
\emph{Skate condition}. The blade cannot move normal to itself, only tangentially to its own direction at a given point. 
In other words,  the velocity $\dot{\br}$ must be parallel to the direction of the skate  and lie in the plane of the ice. In the skate's frame, this condition is 
\begin{equation}
\label{non_holonomic_constr} 
\aaa{\bY ,\bE_1 \times \bG} = 0\, , 
\end{equation} 
which is the non-holonomic constraint. 
\end{enumerate}
The  constraints \eqref{no_tilt} and \eqref{no_lift}  can be expressed as functions of coordinates only (position and orientation of the skater) and so are called \emph{holonomic} constraints. The third condition cannot be expressed as the functions of coordinates only and defines a \emph{non-holonomic} constraint linear in velocities. 
In what follows also neglect all friction forces in the gliding motion of the skate (i.e. the motion along its direction) and friction forces associated with the rotation of the skate on ice. 

\subsection{ Equations of Motion.}
The equations of motion are computed as the balance of linear and angular momenta in the skate's frame using methods of non-holonomic mechanics (Lagrange-d'Alembert's principle) for the Lagrangian \eqref{Lagr_body}, holonomic constraints  \eqref{no_tilt} and \eqref{no_lift},  and non-holonomic constraint \eqref{non_holonomic_constr} \cite{Bloch2003,Ho2011_pII,FeKo1995}.  
Introduce the variations $\bS := (\Lambda^T\delta\Lambda)^\vee$ and $\bp := \Lambda^T\delta\br$, with $(\bS,\bp) \in \mathfrak{se}(3)$, the Lie algebra of $SE(3)$. We have defined $\mathbf{a}=a^\vee$ to be the mapping between $3 \times 3$ antisymmetric matrices $a$ and vectors $\mathbf{a} \in \mathbb{R}^3$ given by the inverse of the hat map: 
$\widehat{a}_{i j}=-\epsilon_{i j k} a_k$, with $\epsilon_{i j k}$ denoting the absolutely antisymmetric Levi-Civita tensor. The variations of the holonomic constrains \eqref{no_tilt} and \eqref{no_lift} yield
\begin{equation}
\label{delta_var} 
\delta\aaa{\bE_1,\bG} = \aaa{\bS,\bE_1\times\bG} \, , 
\quad 
\delta\aaa{\bR,\bG} = \aaa{\bp,\bG} \, . 
\end{equation}
We apply the Lagrange-d'Alembert critical action principle 
\begin{equation} 
\label{LdA-skater} 
0=\delta\int[L(\bO,\bG,\bY) + \kappa\aaa{\bE_1,\bG} + \lambda\aaa{\bR,\bG} ]dt + \mu\int\aaa{\bp,\bE_1\times\bG}dt 
\end{equation} 
 with variations of $\bO$, $\bG$ and $\bY$ satisfying
\begin{equation}
\delta\bO = \dot{\bS} + \bO\times\bS, \ \delta\bG = \bG\times\bS, \ \delta\bY = \dot{\bp} + \bO\times\bp + \bY\times\bS \, , 
\end{equation}	
with $\kappa$ and $\lambda$ enforcing \eqref{no_tilt} and \eqref{no_lift}, respectively, and $\mu$ enforcing the condition on variations $\aaa{\bp, \mathbf{E}_1 \times \bG}=0$ coming from the Lagrange-d'Alembert's principle for non-holonomic constraints. The Lagrange multipliers are proportional to the corresponding magnitudes of reaction forces created by the constraints. 
The terms proportional to $\bS$ and $\bp$ give, respectively, the balances of angular and linear momenta. The same equations can be derived by the Hamilton-Pontryagin principle as outlined in \cite{Ho2011_pII}, Euler-Poincar\'e Suslov's principle \cite{FeKo1995} or other methods of non-holonomic mechanics.  The equations for the angular and linear momentum are, respectively, given by 
\begin{equation} 
\left\{ 
\begin{aligned} 
\label{ang_lin_mom} 
\left( \frac{d}{d t} + \bO \times \right)&  \pp{L}{\bO} + \bG \times \pp{L}{\bG} + \bY \times \pp{L}{\bY}  =  
\kappa \big( \mathbf{E}_1 \times \bG \big) 
\\ 
\left( \frac{d}{d t} + \bO \times \right)&  \pp{L}{\bY} = \lambda \bG + \mu  \big( \mathbf{E}_1 \times \bG \big)
\end{aligned} 
\right. 
\end{equation} 
Using the Lagrangian \eqref{Lagr_body} for the case of  a static skater (\emph{i.e.}, a skater with non-articulated limbs) gives the equations of motion:
\begin{equation} 
\left\{ 
\begin{aligned} 
& \left( \frac{d}{d t} + \bO \times \right)  \bPi - mg \bG \times \bA + \bY \times \mbP  =  
\kappa \big( \mathbf{E}_1 \times \bG \big)  
\\ 
& \left( \frac{d}{d t} + \bO \times \right) \mbP = \lambda \bG + \mu  \big( \mathbf{E}_1 \times \bG \big)
\\ 
&\quad \mbP =  m(\bY +\bO \times \bA) \,, \quad \bPi= \mathbb{I} \bO + \bA \times \mbP 
\\ 
&\quad  \dot \bG = - \bO \times \bG \, . 
\end{aligned} 
\right. 
\label{eqs_of_motion}
\end{equation} 
Here, $\bPi$ and $\mbP$ are expressions for the angular and linear momenta, respectively, expressed in the skate's frame. The first equation of \eqref{eqs_of_motion} is the balance of angular momentum in the frame of the skate and the second equation is the balance of linear momentum. 

Equations \eqref{eqs_of_motion}, taken together with the constraints \eqref{no_tilt}, \eqref{no_lift} and \eqref{non_holonomic_constr}, form the complete and closed description of the system. Once the solution $(\bO, \bY, \bG)$ is found from \eqref{eqs_of_motion}, the position of the skate on ice $\br$ and the $3 \times 3$ matrix $\Lambda$ describing the rotation of the skate with respect to the spatial frame, can be computed using 
the vector $\bO=(\Omega_1,\Omega_2,\Omega_3)$ and $\bY$ as $\dot \Lambda = \Lambda \widehat{\Omega}$ and $\dot \br= \Lambda \bY $, where, as before,  $\widehat{\Omega}_{ij} = - \epsilon_{i jk} \Omega_k$.

Taking the time derivative of the constraint $\aaa{\br,\be_3} = 0$ and expressing it in the skate's frame gives $\aaa{\bY,\bG} = 0$ and together with \eqref{non_holonomic_constr} we can write 
 \begin{equation} 
 \label{Y_constr} 
 \bY = v(t) \mathbf{E}_1.
 \end{equation} 
  Thus, the velocity is only directed along the skate and  $v(t)$ is the speed of the skate at a given point.  
Note that this condition does not mean that the skate is moving in a straight line, since $\mathbf{E}_1$ is  rotating when viewed from the spatial frame. 

While the methods used to derive the equations \eqref{eqs_of_motion} are well-established, these equations, as far as we are aware, are new and have not been derived in the literature before. However, the most interesting part of the problem is not the equations themselves, but the surprising integrability and rich chaotic behavior exhibited by the solutions to these equations, which will be the focus of the remainder of this paper. As it turns out, the integrability vs chaotic behavior is dependent only  on the parameters of the skater, and not on the initial conditions. 


We shall also briefly remark here that the steady states of \eqref{eqs_of_motion}, when they exist, are given by circular or straight motion with the constant angular and linear velocities. These steady states are, in our opinion, not particularly interesting compared to the dynamics of the system, and we skip it for the sake of brevity. We shall thus focus on the fully nonlinear dynamics for the remainder of the paper. As it turns out, for further analytical and numerical progress it is useful to express the equations of motion \eqref{eqs_of_motion} in an interesting basis described below. 

\subsection{ Hybrid frames }
\label{sec:hybrid_frame} 
We transform the system to a basis  $\{ \balpha_1,\balpha_2,\balpha_3\} = \{ \mathbf{E}_1, \bG, \mathbf{E}_1 \times \bG \}$. Note that $\balpha_1 \times \balpha_2 = \balpha_3$, so the basis is orthonormal. Technically speaking, this basis mixes the vectors from spatial ($\bG$) and skate's ($\mathbf{E}_1$) variables, which is rather unusual. We shall call that basis a \emph{hybrid} frame.  

First, note that according to a time derivative of the constraint $\aaa{\mathbf{E}_1, \bG}=0$, we have $\aaa{\bO ,\balpha_3}=0$, so 
\begin{equation} 
\label{omega_expr} 
\bO= \Omega_1 \balpha_1 + \Omega_2 \balpha_2 \, , \quad \Omega_3 := \aaa{\bO,\balpha_3} =0 \, . 
\end{equation} 
In order to compute the equations of motion \eqref{eqs_of_motion} in the $\{ \al_1,  \al_2,  \al_3 \}$ frame, we also need to compute the time derivatives of the basis vectors 
\begin{equation} 
\label{dot_alpha} 
\begin{aligned} 
\dot \balpha_1 & =0 \, , 
\\
\dot \balpha_2 & = \bG \times \bO = \balpha_2 \times \balpha_1 \Omega_1 = - \balpha_3 \Omega_1 
\\ 
\dot \balpha_3 &= \mathbf{E}_1 \times ( \bG \times \bO) = - \balpha_1 \times \balpha_3 \Omega_1 = \balpha_2 \Omega_1 
\end{aligned}
\end{equation} 
We define the torques $\bT$ and forces $\bN$ in the $\{ \al_1,  \al_2,  \al_3 \}$ frame: 
\begin{equation} 
\label{eq_T_N} 
\begin{aligned} 
\bT & = -\Omega_1\Omega_2 \bI \al_3 + \bO\times\bI\bO + mg\bA\times\al_2 \, , 
\\
\bN & = m(\bO\times\bY   +\Omega_1\Omega_2\bA\times\al_3  + \bO\times\bA)\, . 
\end{aligned}
\end{equation} 
At each time step, we have to solve for six variables $\bX=(\dot \Omega_1, \dot \Omega_2, \dot v, \lambda, \mu, \kappa)$ through the linear system 
 $\mathbb{M}\bX = (\bT,\bN)$. Alternatively, we artificially introduce the variable $\dot{\Om}_3 = 0$ and at each time step, we have to solve for the variables $\bX=(\dot \Omega_1, \dot \Omega_2, \dot \Omega_3, \dot v, \lambda, \mu, \kappa)$ through the system 
\begin{equation} 
\label{lin_system_derivs}
\left(
\begin{array}{ccccc} 
\big[  \mathbb{I}_\alpha \big] &    \begin{array}{c}0 \\ 0 \\ 0 \end{array} & 
			 \begin{array}{c} -A_{\alpha3} \\ 0 \\ A_{\alpha1} \end{array}    & 
					 \begin{array}{c} A_{\alpha2} \\-A_{\alpha1} \\ 0 \end{array}   & 
						 \begin{array}{c} 0 \\ 0 \\ -1 \end{array}  
\\
m \big[ \widehat{\bA}_\alpha  \big] &     \begin{array}{c}-m \\ 0 \\ 0 \end{array}   & 
			  \begin{array}{c} 0  \\ 1 \\ 0 \end{array}    & 
					 \begin{array}{c} 0  \\ 0 \\ 1 \end{array}    & 
						 \begin{array}{c} 0 \\ 0 \\ 0 \end{array}  
\\ 
\begin{array}{ccc} 0 & 0 & 1 
\end{array} 
& 0 & 0 & 0 & 0 
\end{array} 
\right) \mathbf{\bX} = 
\left( 
\begin{array}{c} 
-T_1
\\
-T_2
\\
-T_3
\\
N_1 
\\ 
N_2
\\
N_3
\\
0 
\end{array} 
\right) 
\end{equation} where $(\bI_{\alpha})_{i,j}  = \aaa{\al_i,\bI\al_j}$, $(A_\alpha)_i = \aaa{\bA,\al_i}$ and $\big[\widehat{\bA}_\alpha\big]$ is a $3 \times 3$ matrix with $(i,j)$ component computed as 
 \[ 
\big[\widehat{\bA}_\alpha\big]_{ij}=\aaa{\al_i,\widehat{\bA} \al_j}= \aaa{\al_j \times \al_i, \bA}=- \epsilon_{i j k} \aaa{\al_k, \bA}
\]  

Note that the equation for $\Omega_3=\aaa{\bO, \mathbf{E}_1 \times \bG}$ is trivial, $\Omega_3=0$, and $\bI_{\alpha}$ is the moment of inertia computed in the hybrid coordinate system $\{ \al_1, \al_2, \al_3 \}$. 
The equation \eqref{lin_system_derivs} is a system of $7$ equations for $7$ unknowns $\bX$. One can prove  that the $7 \times 7$ matrix in \eqref{lin_system_derivs} is non-singular through a rather tedious, but direct computation of the determinant of that matrix. Thus, a unique solution for $\bX$ in \eqref{lin_system_derivs} can always be found for an arbitrary right-hand side of that equation. 

 For simplicity of calculations, we shall assume that the skater's main axes of inertia  aligned with 
the basis vectors $\{ \mathbf{E}_1, \mathbf{E}_2, \mathbf{E}_3\}$. This assumption is satisfied quite well for a typical position of skater's body used for long gliding motions analyzed here.

\section{Extra constants of motion and conditions for integrability.} 
\label{sec:integrable} 
A general principle of non-holonomic mechanics \cite{Bloch2003} states that the system described here conserves the total energy, calculated as the kinetic plus potential energy. When the skater does not move the parts of the body, the conservation of energy reads: 
\begin{equation}
E \!=\!\frac{1}{2}\!\aaa{\bI\bO,\bO} +  \frac{1}{2}m \norm{\bY +  \bO \times \bA }^2 +mg \aaa{\bA , \bG} \!=\!{\rm const}. 
\label{E_def}
\end{equation} 
 The conservation of energy is normally the only constant of motion one can expect from the system as complex as \eqref{eqs_of_motion}. However, in our case, when the center of mass has no component in the direction of the skate, \emph{i.e.}, $A_1=\aaa{\bA, \mathbf{E}_1}=0$, highly complex equations of motion for the skater \eqref{eqs_of_motion} allow two additional constants (integrals) of motion.
 These additional constants of motion yield integrability of the system and provide  a complete solution of the equations governing the motion of the skater. 

We will now derive  a necessary and sufficient condition for the integrals of motion of that type to exist which turns out to be $A_1=0$.

Before we proceed, it is useful to consider the symmetry of the system. One can see that the Lagrangian \eqref{Lagr_body} and the constraints \eqref{no_tilt}, \eqref{no_lift} and \eqref{non_holonomic_constr} are left-invariant with respect to the rotations and translations along the ice, \emph{i.e.}, the group $SE(2)$. If there is no gravity, \emph{i.e.} $g=0$ in \eqref{Lagr_body}, there is an additional symmetry of rotations about the axis of the blade, so the symmetry group is $SE(2) \times S^1$. The absence of gravity is an unphysical situation, we consider it here only for completeness of the exposition, as it is useful for getting additional mathematical insights into the system.

For the case $g=0$ giving the largest possible symmetry $SE(2) \times S^1$, the relevant components of momenta are the projection of the linear momentum on the axis of the blade $ \aaa{\mathbf{E}_1, \bP}$, the projection of the angular momentum on the vertical axis $ \aaa{\bG, \bPi}$ and the projection of angular momentum on the axis of the blade $ \aaa{\mathbf{E}_1, \bPi}$. These components of momenta are obtained, respectively,  from the linear translations along the blade, rotations about the vertical axis and rotations about the blade's axis when there is no gravity, and are called non-holonomic momenta \cite{BlKrMaMu1996,Bloch2003}. Incidentally, in our case, these quantities are also obtained as the momenta corresponding to velocities $(v, \Omega_1, \Omega_2)$ in the following sense. Define the 
\emph{constrained Lagrangian} which is obtained by substitution of the constrained velocities \eqref{Y_constr} and 
\eqref{omega_expr} into the Lagrangian \eqref{Lagr_body}: 
\begin{equation} 
\label{lagr_constr}
L_c(v, \Omega_1, \Omega_2, \theta) =   L \quad  \mbox{with substitution} \quad  \bO=\Omega_1 \mathbf{E}_1 + \Omega_2 \bG, \, \bY=v\mathbf{E}_1 \, . 
\end{equation} 
Next, take the derivatives of $L_c$ with respect to velocities $v, \Omega_1,\Omega_2$: 
\begin{equation} 
\label{non_hol_mom} 
p_v\!=\!\pp{L_c}{v} \!=\! \aaa{\mathbf{E}_1, \bP} \, , \quad 
p_{\Omega_1}\!=\!\pp{L_c}{\Omega_1}  \!=\! \aaa{\mathbf{E}_1, \bPi} \, , \quad 
p_{\Omega_2}\!=\!\pp{L_c}{\Omega_2} \!=\! \aaa{\bG, \bPi} 
\end{equation} 
Notice that the time derivatives of the quantities $(p_v, p_{\Omega_1}, p_{\Omega_2})$ defined above do not contain the Lagrange multipliers according to the equations of motion \eqref{eqs_of_motion}.

From the definition of non-holonomic momenta \eqref{non_hol_mom}, the most general integral of motion linear in momenta is thus of the form: 
\begin{equation} 
\label{C_gen} 
\begin{aligned} 
C&=\alpha(\theta) p_v + \beta(\theta) p_{\Omega_2} + \gamma(\theta) p_{\Omega_1} 
\\& = \alpha(\theta)  \aaa{\mathbf{E}_1 ,  \bP } + \beta(\theta) \aaa{ \bG , \bPi } + \gamma(\theta) \aaa{ \mathbf{E}_1,\bPi }\, . 
\end{aligned} 
\end{equation} 
Existence of integrals of this type have been investigated both using the Hamiltonian description, under the name of (horizontal) \emph{gauge momenta} \cite{BaGrMa1996,FaGiSa2008,FaGiSa2012,BaSa2016,GaNaMo2018}, and in the Lagrangian description in \cite{BlKrMaMu1996,Ze2003,BlMaZe2009}. We will follow the Lagrangian description in this paper, although the Hamiltonian description of this problem is also of interest and will be considered in later work. 

Our goal is to prove that integrals given by \eqref{C_gen} exists if and only if $A_1=0$. Moreover, we shall prove that there are exactly two independent constants of motion of that type, which is sufficient for integrability. As it turns out, the constants of motion are also computed explicitly in elementary functions, which is quite unusual. We shall also see that $\gamma=0$ for any non-trivial solutions of \eqref{C_gen}, even if $g=0$. 

We are now ready to prove the following 
 \color{black}

 \begin{theorem}[{\em On the integrability of equations using constants of motion linear in momenta}]
 {\rm Let $\mathbb{I}$ be diagonal: $\mathbb{I}={\rm diag} (I_1,I_2,I_3)$. The equations of motion \eqref{eqs_of_motion} are integrable due to the presence of constants of motion linear in momenta \eqref{non_hol_mom}  if and only if $A_1=0$. Moreover, these integrals of motion, when they exist, can be computed explicitly in terms of elementary functions.  
 }
 \end{theorem}

\begin{proof} 
{\em Part A (if)} This part can be proved by direct inspection of equations in $\{ \al_1,\al_2,\al_3\}$ frame. Suppose $A_1=0$. The key to the calculation is to notice by direct inspection that the equations for $\Omega_2= \aaa{\bO,\bG}$ and $v$ for $A_1=0$ are of the form 
\begin{equation} 
\label{A1_redux}
\dot{\Omega}_2=\Omega_1 \Omega_2 f(\theta) \, , \quad \dot v= \Omega_1 \Omega_2 g(\theta) \, ,  
\end{equation} 
where $f(\theta)$ and $g(\theta)$ are some given functions of the inclination angle $\theta$. Remembering that $\dot \theta=\Omega_1$ and using the exact expression for $f(\theta)$ we get the conserved quantity 
\begin{equation}
\label{J_1} 
J_1 = \Omega_2 \aaa{\bG,\bI\bG} =\aaa{\bO, \bG}  (I_2 \sin^2 \theta + I_3 \cos^2 \theta)\, . 
\end{equation}
Furthermore, setting $\Omega_2=J_1/F(\theta)$  in the second equation of \eqref{A1_redux} as $\dot v =J_1 \dot \theta  g(\theta)/F(\theta)$, we get the second conservation law 
\rem{ 
\begin{equation} 
\label{J_2}
J_2=v+ J_1 \int^\theta  \frac{g(s)}{F(s)} \mbox{d} s={\rm const}. 
\end{equation} 
} 
\begin{equation}
\label{J_2} 
J_2 = 
\left\{ 
\begin{aligned}
    & v + 2\Om_2\aaa{\bA,\bE_1\times\bG}& \text{if } I_2=I_3 \\
    & v +   \Om_2\aaa{\bA,\bE_1\times\bG} 
    -\frac{J_1A_2}{\sqrt{I_2|\Delta I |}}\arctanh(\sqrt{\frac{|\Delta I |}{I_2}}\Gamma_3)  
    \\ 
    & \qquad +\frac{J_1A_3}{\sqrt{I_3|\Delta I |}}\arctan(\sqrt{\frac{|\Delta I |}{I_3}}\Gamma_2)          & \text{if} \ I_2 >I_3  \\
   & v  + \Om_2\aaa{\bA,\bE_1\times\bG} 
   -\frac{J_1A_2}{\sqrt{I_2|\Delta I |}}\arctan(\sqrt{\frac{|\Delta I |}{I_2}}\Gamma_3) 
   \\ 
   & \qquad  +\frac{J_1A_3}{\sqrt{I_3|\Delta I |}}\arctanh(\sqrt{\frac{|\Delta I |}{I_3}}\Gamma_2)              & \text{if} \ I_3 > I_2 
\end{aligned} 
   \right. 
\end{equation}
For the case of a realistic figure skater, $I_2$, the moment of inertia about the axis going through the side of the body, is always larger than $I_3$, the moment of inertia about the axis going through the torso up through the head, so we only need to consider the second case in \eqref{J_2} when treating practical applications.

{\em Part B) (only if) } Our goal is to prove that constants of motion given by \eqref{C_gen} exist only if $A_1=0$. 
We would like to choose the functions $\alpha(\theta)$, $\beta(\theta)$ and $\gamma(\theta)$ in \eqref{C_gen} such that $\dot C=0$. For shortness of calculations, let us introduce some notation. We denote 
\begin{equation} 
\label{phi_def}
\varphi(\theta) =  \aaa{\mathbf{E}_1 \times \bG,\bA} = - A_2 \cos \theta + A_3 \sin \theta
\end{equation} 
and notice that 
\begin{equation} 
\label{phi_prime} 
\varphi'(\theta) = A_2 \sin \theta+A_3 \cos \theta = \aaa{\bG,\bA }\, . 
\end{equation} 
We also note that $\mathbb{I} \bO = I_1 \Omega_1 \mathbf{E}_1 + \Omega_2 \mathbb{I} \bG$. 
The time derivative of $C$ is given by 
\begin{equation} 
\begin{aligned} 
\dot C & = \alpha'(\theta) \Omega_1\aaa{ \mathbf{E}_1 , \bP } + \alpha(\theta) \aaa{\mathbf{E}_1 , \dot \bP }
\\
& \quad + \beta'(\theta) \Omega_1  \aaa{ \bG, \bPi }+ \beta(\theta) \aaa{ \bG , \dot \bPi }- \beta(\theta) \aaa{(\bO \times \bG),  \bPi }
\\
& \quad + \gamma'(\theta) \Omega_1 \aaa{\mathbf{E}_1 , \bPi }+ \gamma(\theta) \aaa{\mathbf{E}_1 , \dot \bPi}  \, . 
\end{aligned} 
\label{Cdot} 
\end{equation} 
Computing
 the quantities $\aaa{\mathbf{E}_1, \bP}$, $\aaa{\mathbf{E}_1, \bPi}$,  $\aaa{\bG, \bPi}$ and $\aaa{\dot \bG,  \bPi}=\aaa{- \bO \times \bG, \bPi}$ from definition, and 
$\aaa{\mathbf{E}_1, \dot \bP}$, $\aaa{\mathbf{E}_1, \dot \bPi}$,  and $\aaa{\bG, \dot \bPi}$ from the equations of motion \eqref{eqs_of_motion}, we obtain 
a quadratic polynomial in terms of velocities $\Omega_1$, $\Omega_2$ and $v$. There are only five  combinations of these variables encountered in \eqref{Cdot}, namely $\Omega_1^2$, $\Omega_1 \Omega_2$, $\Omega_2^2$, $v \Omega_1$ and $v \Omega_2$, and there is also a component that is independent of $\Omega_1, \Omega_2$ and $v$ coming from the gravity term. For $\dot C$  to vanish identically, the coefficients of these monomials in velocities must vanish. After some extensive algebra, we get six equations for three functions $\alpha(\theta)$, $\beta(\theta)$ and $\gamma(\theta)$: 
\begin{equation} 
\label{alpha_beta_gamma_eqs}
\hspace{-1mm} 
\begin{aligned} 
\Omega_1^2: \quad  & - \beta' m A_1 \varphi'+ \gamma' \widetilde{I}_1=0, \quad 
 \widetilde{I}_1:= I_1 + m (A_2^2 + A_3^2)   
\\ 
\Omega_1 \Omega_2: \quad   & \beta' \left(  \aaa{\mathbb{I} \bG , \bG} + m \varphi^2  +mA_1^2 \right) 
\\ 
& \quad - mA_1(\varphi'\gamma' -\varphi\gamma) + m 
\left( - \alpha \varphi'+ \alpha' \varphi \right)  =0 
\\ 
\Omega_2^2: \quad & - \gamma \left(\aaa{ \mathbb{I} \bG ,\bE_1\times \bG} - m \varphi \varphi' \right) + \alpha m A_1 =0 
\\ 
v \Omega_1: \quad &\beta' \varphi + \beta \varphi' + \alpha' =0 
\\ 
v \Omega_2: \quad & -\beta A_1 + \gamma \varphi' =0 
\\ 
(\mbox{gravity}): \quad  & m g \gamma \varphi =0 
\end{aligned} 
\end{equation} 
These six equations for three unknowns $\alpha$, $\beta$ and $\gamma$ are, in general, not compatible. Let us consider the solution in more details, as it turns out, the system is solvable if and only of $A_1=0$, independent of the presence of the gravity term. 

If the gravity term, \emph{i.e.}, the last equation of \eqref{alpha_beta_gamma_eqs}, is present, we have to set $\gamma=0$  leading  to either $A_1=0$ or $ \beta = 0$ from the fifth equation of that system. Suppose we choose $A_1 \neq 0$, then $\beta=0$ and also $\alpha=0$ from the third equation of that system, so $\alpha=\beta=\gamma=0$ and $C=0$. So if gravity is present, we must consider $A_1=0$ for nontrivial solution for $C$ to exist. 

If the gravity is absent, \emph{i.e.}, $g=0$,  the last equation of \eqref{alpha_beta_gamma_eqs} is identically satisfied.  
From the fourth equation of \eqref{alpha_beta_gamma_eqs}, we obtain 
\begin{equation} 
\label{v_coeff} 
\alpha + \beta \varphi = K={\rm const} \, . 
\end{equation} 
Suppose $A_1 \neq 0$. Expressing $\beta$ from the fifth equation of \eqref{alpha_beta_gamma_eqs} as $\beta=\varphi' \gamma/A_1$ and substituting into the third equation of the system, with the use of \eqref{v_coeff} gives a linear \emph{algebraic} equation for $\gamma(\theta)$  which can be solved exactly, with explicit solution $\gamma=\gamma(\theta)$. By a similar procedure, using $\beta=\varphi' \gamma/A_1$ in the first equation of the system  gives a linear ODE for $\gamma$ which is incompatible with the solution $\gamma(\theta)$ obtained from the third equation. Thus, the case $A_1 \neq 0$ does not yield any nontrivial solution for $C$. 

Therefore, we must have $A_1=0$ for equations \eqref{alpha_beta_gamma_eqs} to have a non-trivial solution, whether or not the gravity is present. Let us show that this case is compatible with the calculation of part A) and there are exactly two independent integrals.  From the fifth equation of \eqref{alpha_beta_gamma_eqs}, $\gamma=0$, and the last equation of \eqref{alpha_beta_gamma_eqs} is satisfied independent of the value of $g$. For $A_1=0$ and $\gamma=0$, the first and third equations of \eqref{alpha_beta_gamma_eqs} are also identically satisfied. The system reduces to just two equations for $\alpha$ and $\beta$ obtained from the second and fourth equations of the system for $\gamma=0$ and $A_1=0$: 
\begin{equation} 
\label{simple_alpha_beta}
\begin{aligned}
&\beta' \left(  \aaa{\mathbb{I} \bG , \bG} + m \varphi^2 \right)  + m 
\left( - \alpha \varphi'+ \alpha' \varphi \right)  =0
\\ 
&\alpha+ \beta \varphi=K={\rm const} 
\end{aligned} 
\end{equation} 
Substitution of $\alpha$ from last equation into the first equation of \eqref{simple_alpha_beta} yields a single equation for $\beta$ involving the parameter $K$: 
\begin{equation} 
\label{beta_sol} 
\beta'    \aaa{\mathbb{I} \bG , \bG}   - mK \varphi' =0 \, , 
\end{equation} 
leading to 
\begin{equation} 
\label{beta_sol2} 
\begin{aligned} 
\beta(\theta) & = \beta_0 + K \beta_1 (\theta), 
\\
\beta_1(\theta) &:= \int^\theta \frac{\varphi'}{\aaa{\mathbb{I} \bG , \bG}  } \mbox{d} u = 
 \int^\theta \frac{A_2 \sin u+A_3 \cos u}{I_2 \sin^2 u+ I_3 \cos^2 u} \mbox{d} u \, . 
\end{aligned} 
\end{equation} 
The solution \eqref{beta_sol} leads to the following expression for $C$ 
\begin{equation} 
\label{C_simplified2} 
\begin{aligned} 
C&=m v  K + m \Omega_2 \left( K + (\beta_0 + K \beta_1 (\theta)   \aaa{\mathbb{I}\bG, \bG} \right) 
\\ & 
= \beta_0 m \Omega_2 \aaa{ \mathbb{I} \bG, \bG} + m K \left[   v +  \Omega_2 \varphi + \Omega_2 \beta_1 (\theta)   \aaa{\mathbb{I}\bG, \bG}  \right] \, . 
\end{aligned} 
\end{equation} 
This expression \eqref{C_simplified2} contains both \eqref{J_1} and \eqref{J_2} derived above. Indeed, 
if $K=0$, and $\beta_0=1$ then $C=m J_1$. If $\beta_0=0$ and $K=1$, then $C=m J_2$. The theorem is proved. 

\end{proof} 

The constants of motion \eqref{J_1} and \eqref{J_2} are sufficient to completely solve the problem no matter how complex the apparent motion of the skate may be. From the purely geometric point of view, the system is moving in four dimensional space $(v,\Omega_1,\Omega_2,\theta)$. Each of the three constants of motion $E$, $J_1$ and $J_2$ reduces the dimension of the available space by one, so the resulting motion is  along a one-dimensional curve, which must be closed since the conservation energy limits the range of the available motion. 

Moreover, 
we can express $\Omega_1=\dot \theta$ in terms of the constants $(E, J_1, J_2)$ and the tilt angle $\theta$ and thus find a solution $\theta=\theta(t)$, yielding  all other variables, $(\Omega_1, \Omega_2, v)$ and the position of the skate on the ice. However, that solution is algebraically quite complex and hardly informative, so we do not present it here. Instead, we illustrate the motion using a particular example of simulations of equation  \eqref{eqs_of_motion} n the Section~\ref{sec:chaotic} below.

\begin{remark}[{\em On the connection of the basis $\{ \al_1,\al_2,\al_3\}$ and Hamel's frames}] 
{\rm The othonormal set of vectors $\{ \al_1,\al_2,\al_3\}$ selects a basis in both $\bO$ and $\bY$ variables, \emph{i.e.} in the Lie algebra $\mse(3) \simeq \mathbb{R}^3 \times \mathbb{R}^3$. The basis in both the rotational ($\bO$) and translational ($\bY$) parts of $\mse(3)$ is identical. The choice of this basis is quite close to the framework of Hamel's method of quasivelocities \cite{BlMaZe2009}. Unfortunately, even in this basis, we do not achieve a complete removal of the Lagrange mutipliers from the equations for general $\bA$, as is evidenced by \eqref{lin_system_derivs}.

While there are other ways to solve \eqref{eqs_of_motion} numerically, we have found that the use of the hybrid $\{ \al_1,\al_2,\al_3\}$ frame affords the simplest treatment of the constants of motion by providing a specific form of equations of motion \eqref{A1_redux}, and the corresponding derivations of the first integrals \eqref{J_1} and \eqref{J_2}, which seems quite difficult to achieve for alternative choices of the basis.   
If we had chosen $SO(3) \times \mathbb{R}^2$ as the configuration manifold, we could choose the basis in $\bO$ variables as $\{ \al_1,\al_2,\al_3\}$ and in $\bY$ variables as $\{ \al_1,\al_3\}$, which is less symmetric than the representation we have chosen here, but certainly possible to use as well. 
}
\end{remark} 

\begin{remark}[{\em On symmetry considerations and existence of first integrals of the type \eqref{C_gen}}]
{\rm We shall note that sometimes the constants of motion in non-holonomic systems arise from symmetry, although the situation is considerably more complex than the classical case, when any continuous  symmetry of a mechanical system leads to a conserved quantity. This result is known as the Noether theorem and is the reason behind the linear and angular momenta conservation in mechanics, which follow from the symmetries with respect to translations and  rotations \cite{MaRa2013,Ho2011_pII}. 
In non-holonomic mechanics, the situation is considerably more complex, see, for example, \cite{BlKrMaMu1996,Sn1998,FaSa2009,BoMa2015}. The case $g=0$ gives the most striking illustration of the complexity of the problem. Indeed, when $g=0$, there is no potential energy in \eqref{Lagr_body} and the variable $\theta$ does not enter either the Lagrangian or the constraints explicitly, so it is cyclic.  However, the corresponding momentum $p_{\Omega_1}$ defined by \eqref{non_hol_mom} is not conserved, as it would have been in a holonomic system. This effect is due exclusively to the presence of non-holonomic constraints.  It is also quite remarkable that the integrability through the existence of two first integrals of the type \eqref{C_gen} is independent of gravity, \emph{i.e.},  of the fact whether the symmetry group is $SE(2)$ or $SE(2) \times S^1$.  
}
\end{remark}

\rem{ 
We shall note that sometimes the constants of motion in non-holonomic systems arise from symmetry, although the situation is considerably more complex than the classical case, when any continuous  symmetry of a mechanical system leads to a conserved quantity. This result is known as the Noether theorem and is the reason behind the linear and angular momenta conservation in mechanics, which follow from the symmetries with respect to translations and  rotations \cite{MaRa2013,Ho2011_pII}. 
In non-holonomic mechanics, the situation is considerably more complex, see, for example, \cite{Sn1998,FaSa2009,BoMa2015}. In our case, the symmetries that do not change the Lagrangian \eqref{Lagr_body} are rotations about the vertical axis and translations along the ice. In the case of $A_1=\aaa{\bA, \mathbf{E}_1}=0$, rotating the system about the vertical axis going through the center of mass also moves the skate parallel to itself, compatible with the constraint, which is not the case for $A_1 \neq 0$. 
} 

We note that \cite{BlMaZe2009} derive conditions for the existence of integrals linear in non-holonomic momenta such as \eqref{C_gen}, although they do remark that explicit calculation of these integrals is non-trivial even for the simple cases. We have presented the direct analytical calculation in the proof of the Theorem above as it gives the explicit expression for the first integrals, rather than the conditions for their existence. 

To further connect our results with previous works on integrability of non-holonomic systems, 
we remark that \cite{VeVe1988} discusses the integrability of the so-called LR systems on Lie groups, having a right-invariant connection and left-invariant metric, which is important for extensions of this model to more complex physical cases.  Another useful direction is to consider the existence of invariant measure in our system, as discussed in \cite{Ko1985,Ko1988,Ko2002,BlMaZe2009}.   These approaches may be helpful for further investigation of integrability of motion in a general setting, for example, whether either one of the constants of motion $J_1$ and $J_2$ persist for an articulated skater preserving some conditions for integrability, such as $A_1=0$. 

One would expect from experience that an integrable case  would require to the center of mass being aligned with the axis $\mathbf{E}_3$, for any  kind of regular motion to exist. It is thus even more surprising that the integrability described here  exists for all sideways shifts of  the center of mass $\bA$, \emph{i.e.}, all values of $\bA=(0,A_2,A_3)$.

\section{Numerical studies of integrable vs non-integrable case, transition and chaotic behavior} 
\label{sec:chaotic} 
When $A_1 \neq 0$, the energy is still conserved, but the quantities $J_1$ and $J_2$ described by \eqref{J_1} and \eqref{J_2}, respectively, are not conserved and there is no integrability. To contrast the cases of $A_1=0$ and $A_1 \neq 0$, we consistently perform two sets of numerical simulations of equations \eqref{eqs_of_motion}. 
All simulations presented here are performed 
for the values of parameters $m= 50$ kg and moments of inertia being  $I_1 = 15.95$ kg$\cdot m^2$ (rotation axis along the skate), $I_2 = 13.56$ kg$\cdot m^2$ (rotation about the sideways axis), $I_3 = 3.99$  kg$\cdot m^2$ (rotation about the vertical body axis going from the skate to the head). The center of mass is taken to be at $\bA=(A_1,A_2,A_3)$ in the frame of the skate, with $A_2 = 0.12 \, m$  (sideways axis), $A_3 = 0.875\, m $  (vertical body axis) and two cases,  $A_1=0 \, m$ (integrable case) and $A_1=0.1 \, m$   (non-integrable case).  The initial conditions $\Om_1(0) = 0.01s^{-1}$ (rotation about the skate's axis), $\Om_2(0) = 1.25 s^{-1}$ (rotation about the vertical) and $v(0) = 0.5 m/s$.

\begin{figure}[H]
\centering 
\includegraphics[width=0.8\textwidth]{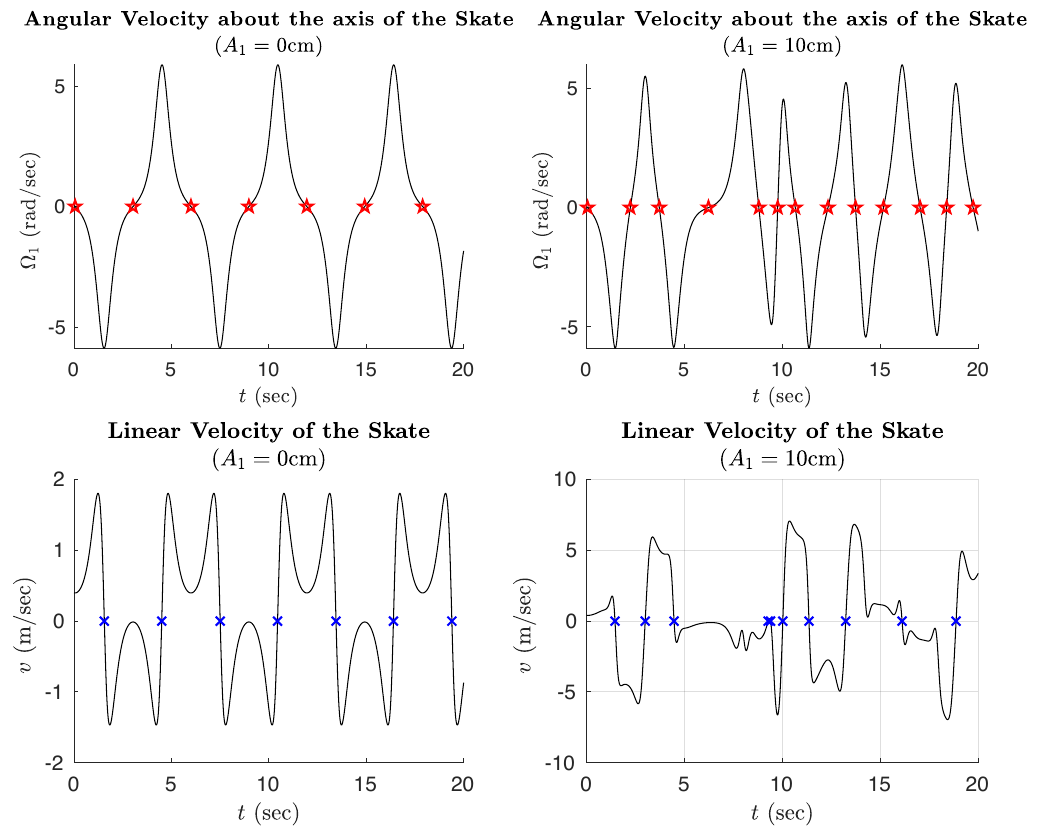} 
\caption{The component $\Omega_1 = \dot \theta$ and $v$ as functions of $t$ for the integrable case $A_1=0$ (left) and non-integrable case $A_1\neq0$ (right). The red stars correspond to zeros in $\Omega_1=\dot \theta$ and blue crosses correspond to zeros in $v(t)$. These symbols relate the corresponding symbols on Figure~\ref{fig:3D_position}, noting the inflection points of the curve and cusps in trajectory. 
 \label{fig:Omega1_v}
}  
\end{figure}

\begin{figure}[h]
\centering 
\includegraphics[width=0.8\textwidth]{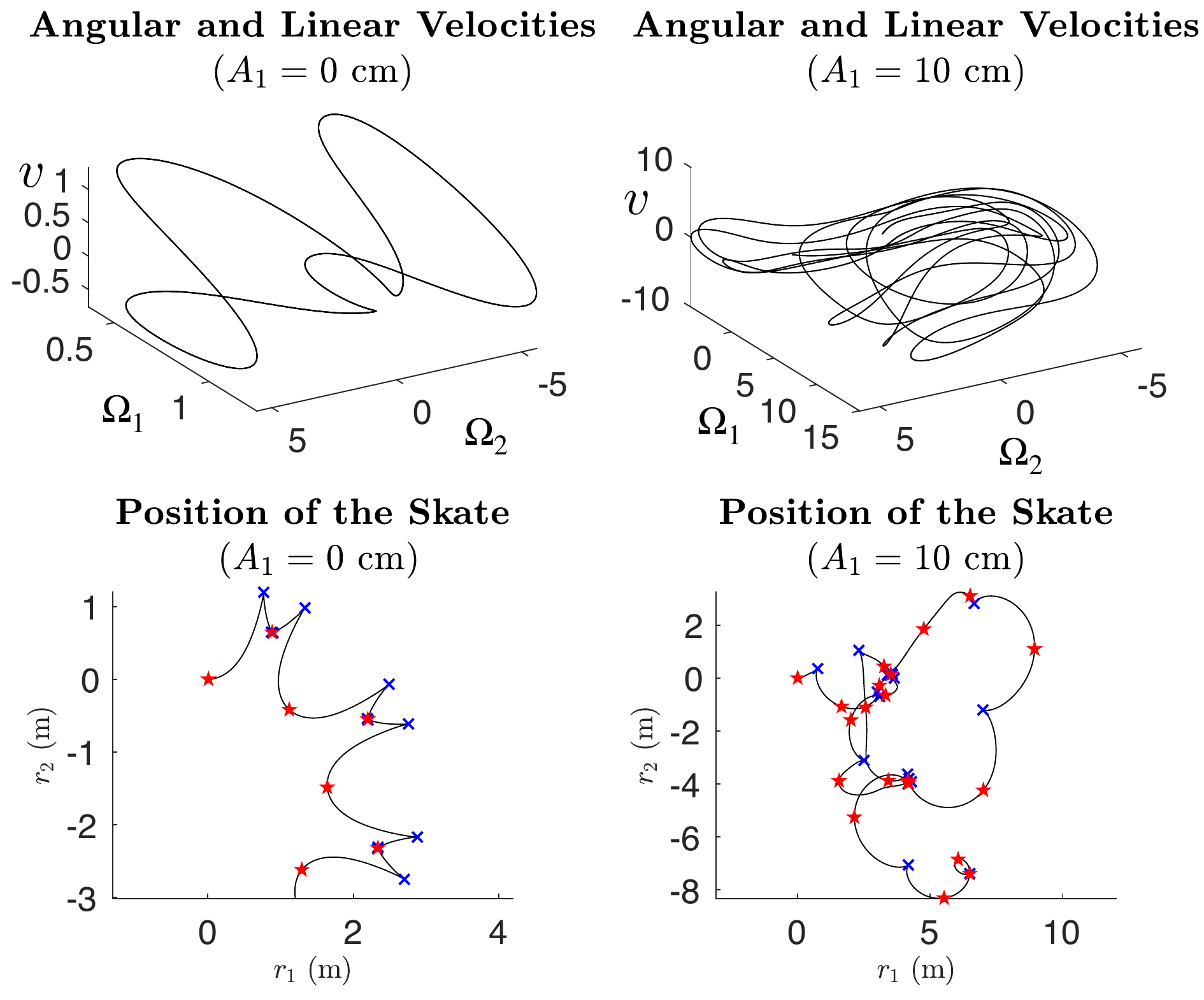} 
\caption{The components $\Omega_1 = \dot \theta$,  $\Omega_2$  and $v$ as functions of $t$ for the integrable case $A_1=0$ (left) and non-integrable case $A_1\neq0$ (right). 
 \label{fig:3D_position}
}  
\end{figure}

Examples of numerical simulations of equation \eqref{eqs_of_motion} are presented on Figure~\ref{fig:Omega1_v}.
Note that in the integrable case  ($A_1=0$) presented on the left panels the motion is clearly periodic whereas on the  case $A_1 \neq 0$ presented on the right panels the motion is apparently irregular. When $A_1=0$, the constants of motion $E$, $J_1$ and $J_2$ are conserved with the expected precision during the simulations, whereas for the non-integrable case $A_1 \neq 0$, only $E$ is conserved whereas $J_1$ and $J_2$ vary considerably. 
These results are further illustrated on Figure~\ref{fig:3D_position} where the figures on the left show a closed curve in the $(\Omega_1= \dot \theta,\Omega_2,v)$ space in the integrable case and a clearly chaotic evolution on the $A_1 \neq 0$ case. Note that all the trajectories presented here, no matter how complex, are obtained for a static skater. 
The bottom panels on that Figure illustrate the trajectory of the skate on ice. The red stars, which mark points of $\Omega_1=\dot \theta=0$, correspond to the inflection points on the trajectory.  The blue crosses correspond to zeros in $v(t)$, yielding  'cusps' where the skate stops and starts going backwards. These cusp points are encountered in real-life figure skating.  We must also emphasize that only the trajectory on ice has a singularity at the cusp, the motion in the space $(\Omega_1,\Omega_2,v,\theta)$ remains regular. 

\begin{figure}[H]
\centering 
\includegraphics[width=0.8\textwidth]{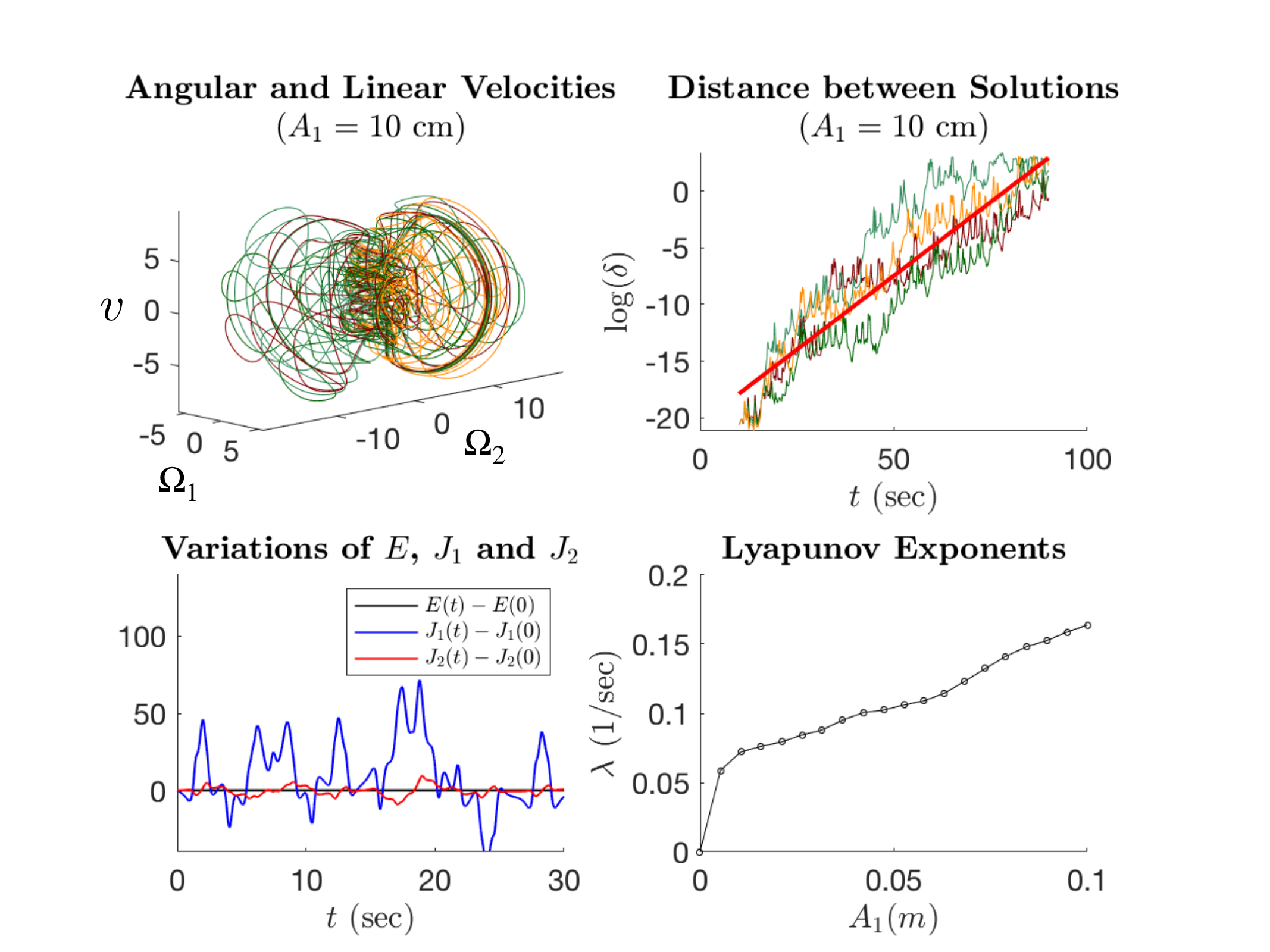} 
\caption{Upper left panel: 4 trajectories plotted in $(\Omega_1,\Omega_2,v)$ space with different colors, depending on the initial conditions having the same energy. Bottom left panel: $E-E(0)$, $J_1-J_1(0)$ and $J_2-J_2(0)$ versus time. In the case $A_1=0$, the plotted quantities vanish up to computational precision. Upper right panel: The  growth of distances between trajectories starting on the same energy surfaces in  $(\theta,\Omega_1,\Omega_2,v)$. Solid red line is the best linear fit describing the growth rate, \emph{i.e.}, Lyapunov exponent for given value of $A_1 \neq 0$. 
Bottom right panel:  Lyapunov exponent $\lambda$  vs $A_1$. 
 \label{fig:chaotic}
}  
\end{figure}

All trajectories with $A_1 \neq 0$ we have tried are chaotic. A set of several trajectories starting with different initial conditions for the same energy and their subsequent analysis, are performed in Figure~\ref{fig:chaotic}.  On the left panel of that Figure, we plot different trajectories for the time $10 sec<t<80 sec$ in the $(\Omega_1,\Omega_2,v)$ space, ignoring the initial time interval $0<t<10 sec$. 

The chaotic nature of the system with $A_1 \neq 0$ is further illustrated in Figure~\ref{fig:chaotic}. The trajectories plotted in $(\Omega_1,\Omega_2,v)$ space in the upper left panel illustrate the complex behavior of the trajectories in that space starting from several initial conditions with the same energy, in the time interval $10<t<80$. The bottom left-hand panel shows that while the energy is conserved, the quantities $J_1$ and $J_2$ are no longer constants.  Two trajectories starting nearby on the same energy surface do diverge and the rate of divergence is approximately exponential, as illustrated on the panel in the upper right-hand corner of that Figure. The rate of divergence of nearby trajectories, also known as the (main) Lyapunov exponent $\lambda$ for a given $A_1$, is computed as the best linear fit to the data $\log \delta(t)$ vs $t$ until saturation $\delta<0.01$ and is shown with the solid red line.  The procedure is repeated and $\lambda$ is measured for values of $A_1$ between $0$ and $0.1$m, as shown on the bottom right  panel. One can see that the Lyapunov's exponent is increasing with $A_1$ and thus the system becomes more chaotic. For $A_1=0$, $\lambda=0$ since the system is integrable. 

\begin{figure}[H]
\centering 
\includegraphics[width=0.8\textwidth]{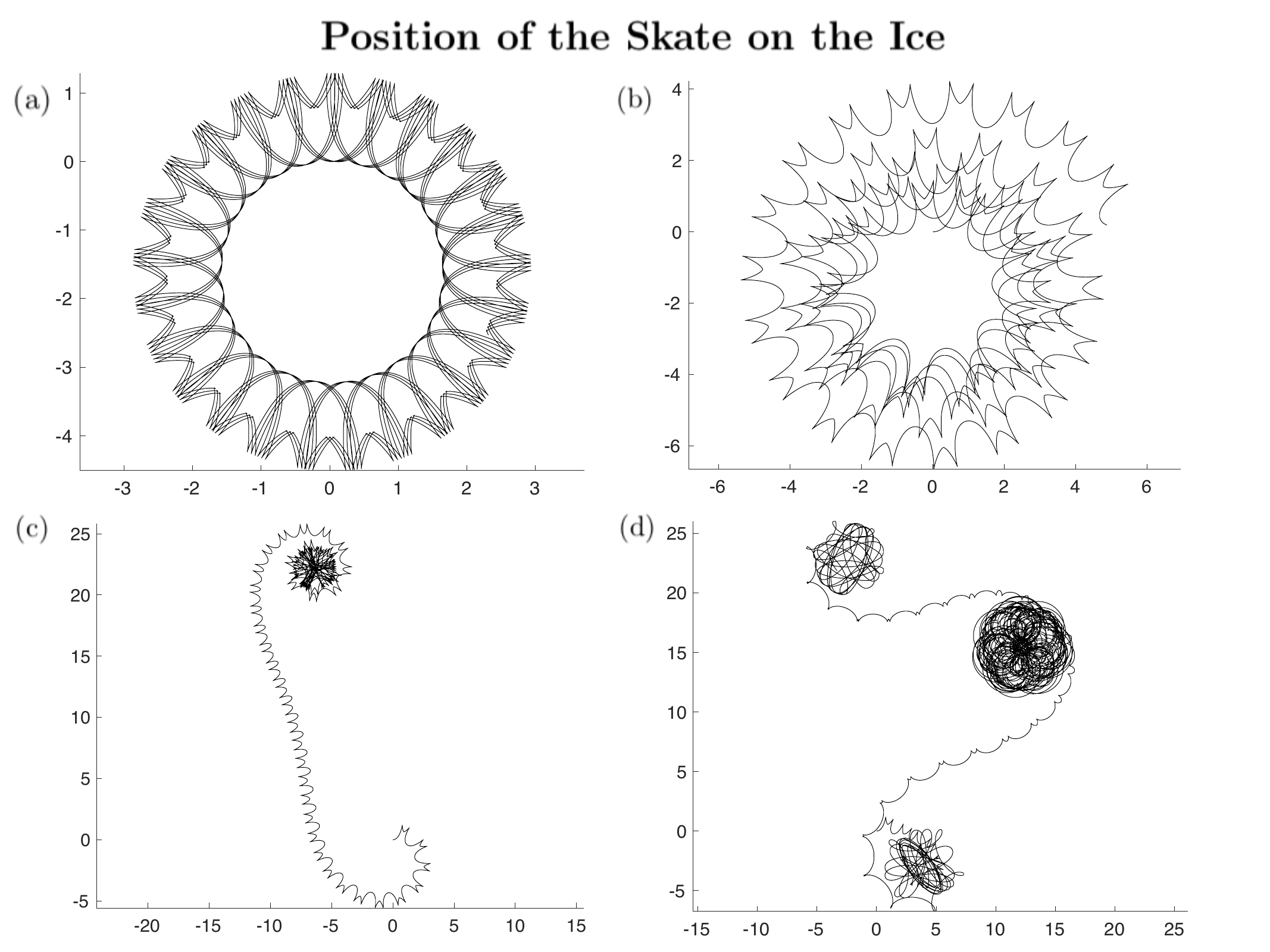} 
\caption{Trajectories on ice of the skater having the same initial conditions for the dynamical variables, and increasing values of $ A_1$. Simulations are run for $0 \leq t \leq 500$ sec. The values of $A_1$ are (left to right and top to bottom):  (a) $A_1=0$m, (b) $A_1 =  10^{-5}$m, (c) $A_1 = 10^{-4}$m, 
{d) $A_1=10^{-3}$m. 
 }
 \label{fig:bifurcation}
}
\end{figure}

One also observes that the bifurcation from integrable case $A_1=0$ to non-integrable case $A_1 \neq 0$ is quite complex and interesting, and we present an 
initial study of that bifurcation. Note that because the system is non-holonomic, we cannot readily make a connection with, for example, the KAM theory for perturbations of integrable Hamiltonian systems \cite{ArKoNe1989}. Figure~\ref{fig:bifurcation} presents the trajectories on ice for increasing values of $A_1$, showing increased chaoticity of the behavior. 

\begin{figure}[H]
\centering 
\includegraphics[width=0.8\textwidth]{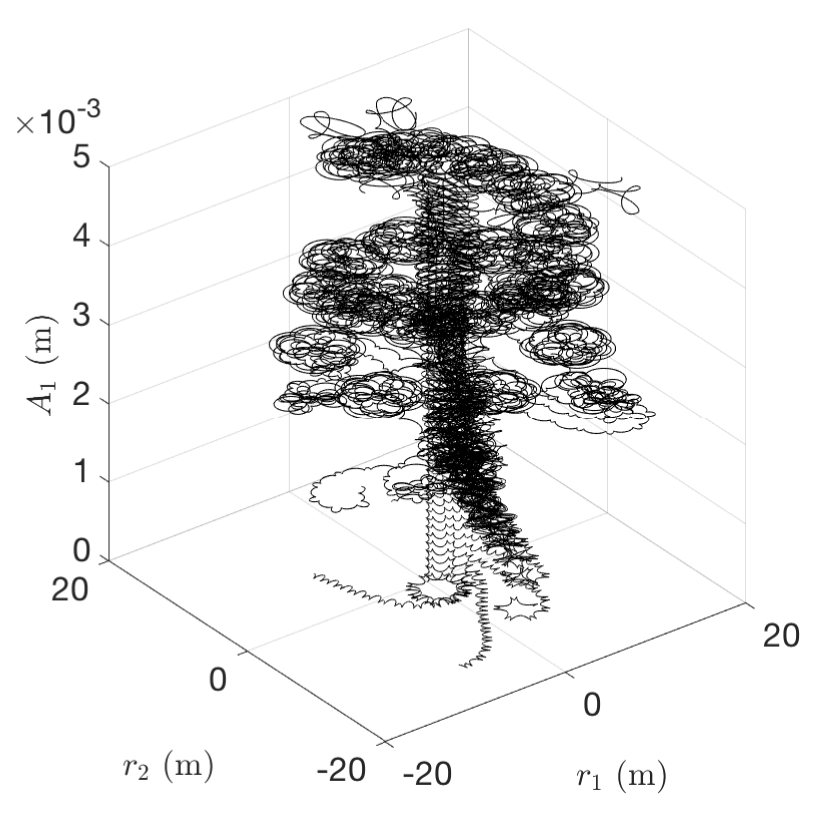} 
\caption{
 \label{fig:bifurcation_all}
 Trajectories on ice for 50 values of $A_1$ equally spaced between $A_1=0$m (integrable case) and $A_1=0.05$m. The trajectories are offset in vertical $(z)$ direction by the 
 corresponding value of $A_1$ so all simulations can be represented simultaneously. 
}  
\end{figure}

To further demonstrate the complexity of this transition from the integrable to chaotic case, Figure~\ref{fig:bifurcation_all} shows ice trajectories starting with identical initial conditions, with  50 values of $A_1$ equally spaced between  $A_1=0$m to $A_1=0.05$m, offset in vertical direction by the value of $A_1$ for easier visual interpretation. 

Because of this highly interesting behavior, we believe that the transition from the integrable to chaotic case, governed by a single parameter $A_1$ presents an interesting and promising case for future numerical and analytical studies. It is possible that the expansion of the motion in terms of unstable periodic orbits \cite{CvChPu1997,Di-etal-2017} can shed more light on the nature of the chaotic system for the case $A_1 \neq 0$.

\section{Conclusions}
\label{sec:conclusions}
 We have derived and analyzed the motion of a three-dimensional static skater based on realistic assumptions on the nature of the skate and its contact with the ice. Surprisingly, one can derive explicit integrals of motion for this problem in quite a general setting, with $A_1=0$ being the necessary and sufficient condition for the integrability due to the presence of additional integrals linear in the momenta. In spite of the apparent complexity of the equations of motion, our system presents a non-trivial example of an integrable non-holonomic system, if $A_1=0$, and chaotic otherwise.The results presented here open the way for further studies, improving the model and making it more realistic. For example, one can consider an articulated skater with a moving center of mass preserving the integrability condition $A_1=0$. It would be interesting to see whether any of the integrals of motion persists, at least for some specific articulation  of the skater.  Another interesting direction for further studies would be the incorporation of friction in the system \eqref{eqs_of_motion}, and whether any friction can preserve the constants of motion derived in \eqref{J_1} and \eqref{J_2}.  Yet another interesting and promising direction would be to to study the diffusion of ice trajectories in the chaotic case, such as presented on Figure~\ref{fig:bifurcation},  using tools of statistical physics, and see whether they correspond to any known physical examples. We shall also note that we are not aware of any general principles for finding the integrals of motion that are nonlinear in velocities or momenta, except for the energy. While their existence in our problem seems unlikely for a general values of skater's parameters, based on the numerical simulations presented here, it is possible that some of the values of $\bA$ and other physical parameters may actually lead to such nonlinear integrals of motion, which will be highly interesting and non-trivial. 
Finally, it is tempting to conjecture that a real-life figure skater intuitively enforces the quantity $A_1$ to be close to zero, and thus keeping the system close to integrability.  It should be feasible to experimentally verify this conjecture using  modern advances in body-tracking technology  \cite{VoBlSch2014}.

\section*{Acknowledgements.} We are grateful to P. Balseiro, A. M. Bloch, F. Fasso, H. Dullin, I. Gabitov, L. Garcia-Naranjo, D. D. Holm, T. Ohsawa, P. Olver, T. S. Ratiu, S. Venkataramani and D. Volchenkov  for enlightening scientific discussions. We want to especially thank D. V. Zenkov for his interest, availability and patience in answering our questions. We are also grateful to M. Hall and J. Hocher for teaching us the intricacies of skating techniques and the differences between hockey and figure skating. We are thankful to M. Hall for providing skating expertise and C. Hansen's graphics processing for Figure~\ref{fig:skater_with_axis}. The research of VP was partially supported by the University of Alberta and NSERC Discovery grant, which also partially supported VG through NSERC USRA program.

\bibliographystyle{unsrt}

\bibliography{FigureSkatingBib}

\end{document}